\documentclass[10pt,twocolumn,notitlepage]{article}
\usepackage{times}
\usepackage{amsfonts}
\usepackage{graphicx}

\begin{document}

\title{Deconstructing the energy landscape: new algorithms for folding heteropolymers}

\author{Veit Elser and Ivan Rankenburg}

\date{Department of Physics,
Cornell University,
Ithaca, NY 14853, USA}

\maketitle

\textbf{
We apply the computational methodology of phase retrieval to the problem of folding heteropolymers. The ground state fold of the polymer is defined by the intersection of two sets in the configuration space of its constituent monomers: a geometrical chain constraint and a threshold constraint on the contact energy. A dynamical system is then defined in terms of the projections to these constraint sets, such that its fixed points solve the set intersection problem. We present results for two off-lattice HP models: one with only rotameric degrees of freedom, and one proposed by Stillinger \textit{et al.}\cite{Stillinger} with flexible bond angles. Our phase retrieval inspired algorithm is competitive with more established algorithms and even finds lower energy folds for one of the longer polymer chains.
}

A favorite metaphor in the field of nonlinear optimization, and computational protein folding in particular, is the \textit{energy landscape}. Energy landscapes have been compared to funnels\cite{funnel}, golf-courses\cite{golf}, and are generally held responsible for all the behavior observed in nature, as well as the challenges faced by simulators. Kinetics simulations are, by their very nature, tied to the topography of the energy landscape and cannot avoid scaling its barriers and languishing in its manifold minima. The outlook for native fold discovery, however, is more optimistic. As we show below, for this problem there are options that escape the confines of the energy landscape and yield significant computational dividends.

Most native fold search strategies are conservative in at least two respects. First, the search is carried out in the same space accessed by the physical degrees of freedom of the protein. Second, the search in this space is carried out quasi-locally, in the sense that every conformation examined is derived from a previously considered conformation by a local modification. There are alternatives to these general guidelines that have proven effective in other fields. For inspiration we turn to the classic problem of \textit{phase retrieval}. 

The naive search space in phase retrieval is superficially equivalent to the space of rotamer configurations, each unknown phase angle $\phi$ corresponding to a dihedral angle on the protein backbone. An important application of phase retrieval is the reconstruction of the electron density in a crystal, given its Fourier amplitudes $F_\mathbf{q}$:
\begin{equation}\label{density}
\rho(\mathbf{r})=\sum_\mathbf{q}F_\mathbf{q} \cos{(\mathbf{q}\cdot\mathbf{r}+\phi_\mathbf{q})}
\end{equation}
The task of the algorithm is to find values for the phases $\phi_\mathbf{q}$ such that the resulting density (\ref{density}) satisfies certain general characteristics (\textit{e.g.} positivity, atomicity) or constraints. To illustrate the idea, we consider a very simple situation where the given amplitudes $F_\mathbf{q}$ are derived from a density known to take only two values, say $\rho =\pm 1$. To implement the binary valued density constraint we could try minimizing a penalty function of the form
\begin{equation}
V=\sum_\mathbf{r} \left(\rho(\mathbf{r})^2-1\right)^2\;,
\end{equation}
where the positions $\mathbf{r}$ fall on a grid determined by the range over which the Fourier vectors $\mathbf{q}$ are sampled. This expression for $V$, an explicit function of the phase variables $\phi_\mathbf{q}$, is a possible energy landscape for the phase retrieval problem. The correct phases are identified by discovering a point on the landscape where the energy realizes the minimum value $V=0$.

Practical phase retrieval algorithms do not minimize an objective function as sketched above\cite{shelx, snb, retriever}. The most successful algorithms do not navigate the barriers and false minima of an energy landscape. Typically, the search performed by these algorithms is carried out in a much larger space (than the space of ``rotamers") and the steps executed are global in character. The example above serves to illustrate the key elements of the search dynamics, called \textit{projections}. There are two projections, both of which act on a density that has been freed of all constraints. In particular, one no longer insists that $\rho$ has the given Fourier amplitudes, that is, the form (\ref{density}) parametrized by phase angles. Instead, one uses the device of a projection $P_A$, which takes an arbitrary input density $\rho$ and returns a minimal modification of $\rho$ where the given Fourier amplitudes have been restored. This can be computed efficiently, by first transforming $\rho$ to Fourier space, making the necessary modification there, and then transforming back. The term ``projection" is derived from the minimality condition, and in the case of $P_A$ corresponds (in Fourier space) to mapping each complex Fourier coefficient to the nearest point on a circle whose radius is given by the corresponding amplitude $F_\mathbf{q}$. The binary constraint on the density values is implemented by another projection, $P_B$, where minimality of the change calls for all positive values to be replaced by $1$, negative values by $-1$. Each of the two projections accomplishes something global, in effect solving half of the problem to completion. The spectrum of modern phase retrieval algorithms arises both from the variety of the kinds of projections used, as well as variations in how they are combined\cite{diffmap}.

\begin{figure}[t]
\begin{center}\includegraphics[width=2in]{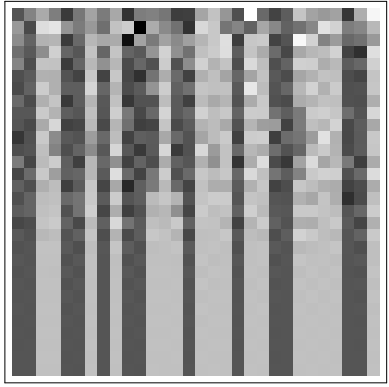}
\end{center}
\begin{center}
\caption{Difference map solution of a phase retrieval problem. Each horizontal row represents one iterate of the one dimensional density, with the initial density at the top and convergence to the binary valued solution at the bottom.}
\end{center}
\end{figure}

Figure 1 shows successive iterates of a particular combination of projections, called the \textit{difference map}\cite{diffmap}, on the phase retrieval problem for a binary valued density. The dynamics is deterministic and the discovery of the solution corresponds to the arrival at a fixed point of the map. Although the number of iterations required by the algorithm depends on the initial density, this number is always much less than the size of an exhaustive search.

We show below that the projection technique can be applied to the protein native fold search problem, and that for simple off-lattice heteropolymer models the results are encouraging. After a brief review of the difference map scheme for combining projections, we examine in detail the two projections that apply to the native fold search. We present results for two HP models,
one with only dihedral degrees of freedom (rotamer model), and a model proposed by Stillinger \textit{et al.}\cite{Stillinger} with variable bond angles (flexible chain model). For the longer chains the projection based algorithm was able to find lower energies than published results\cite{Hsu, Bachmann} obtained by methods that explore the energy landscape.

\section*{Theory and Methods}

\textbf{Difference map algorithm.} The search space is in general a high dimensional Euclidean space $E$. Polymer conformations, for example, are embedded by associating three Cartesian coordinates of $E$ with the position of each monomer in the chain. The goal of the algorithm is to discover one element $x\in A\cap B$, where $A$ and $B$ are subsets of $E$, usually having the character of constraints. In polymer applications, for example, set $A$ might represent all monomer configurations that satisfy the chain constraints (bond lengths, etc.). The constraint sets $A$ and $B$ are assumed to be simple enough that the two projections to these sets, $P_A$ and $P_B$, can be computed efficiently. For example, to compute $y=P_A(x)$, we need to find an element $y\in A$ that minimizes the distance $\|y-x\|$. In difference map applications one may relax the condition that $y\in A$ realizes the true minimum of $\|y-x\|$, although this is usually easy to achieve when $y$ is near enough to $x$ that the constraint can be linearized. In general, the performance of the algorithm is improved by the distance minimizing quality of the projections.

\begin{figure}[t]
\begin{center}\includegraphics[width=3.5in]{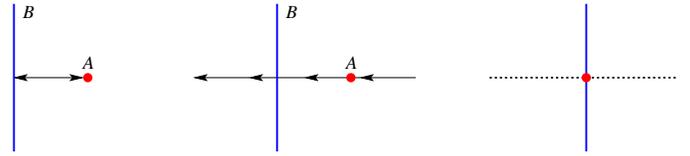}
\end{center}
\begin{center}
\caption{Comparison of alternating projections (left) and difference map iterations (center) in the case of two constraint sets, a point and a line, that do not intersect. The alternating map $P_A\left(P_B(x)\right)$ stagnates on set $A$; iterates of $D(x)$ move uniformly along the axis of nearest separation between $A$ and $B$. When $A$ and $B$ intersect (right), every point in the space locally orthogonal to both constraints is a fixed point of $D(x)$.}
\end{center}
\end{figure}

When the projections are combined in alternating fashion, $x\to P_A\left(P_B(x)\right)$, problems arise when there is a local minimum in the separation of the constraint sets. As shown in Figure 2, this map will then have a fixed point $x^\ast=P_A\left(P_B(x^\ast)\right)$ that lies in $A$ but not $B$. The difference map is a more elaborate combination of projections given by\cite{diffmap}
\begin{eqnarray}
&x\to D(x)=x+\beta\,\Delta(x)\\
&\Delta(x)=P_A\left(f_B(x)\right)-P_B\left(f_A(x)\right)\;,
\end{eqnarray}
where
\begin{eqnarray}
f_A(x)&=&P_A(x)-\beta^{-1}(P_A(x)-x)\;\\
f_B(x)&=&P_B(x)+\beta^{-1}(P_B(x)-x)\;,
\end{eqnarray}
and $\beta$ is a dimensionless parameter. At a fixed point $x^\ast=D(x^\ast)$, we have $\Delta(x^\ast)=0$ and
\begin{equation}
P_A\left(f_B(x^\ast)\right)=P_B\left(f_A(x^\ast)\right)=x_\mathrm{sol}\;.
\end{equation}
This shows that $x_\mathrm{sol}\in A\cap B$, since $x_\mathrm{sol}$ is in the range of both projections. The more straightforward definition $\Delta(x)=P_A(x)-P_B(x)$, which leads to the same conclusion, is not useful because the fixed points $x^\ast$ of $D$ are then unstable. The maps $f_A$ and $f_B$ are tuned to maximize the attraction of the difference map's fixed points\cite{randproj}. When confronted with a near intersection of sets $A$ and $B$, iterates of the difference map move at a uniform rate along the axis of nearest separation, as shown in Figure 2. The step size in the latter situation decreases in proportion to the distance between $A$ and $B$, and the flow degenerates into a space of fixed points when $A$ and $B$ intersect. 

Studies of hard optimization problems, such as phase retrieval, point to the following sequence of events in the difference map solution process. Starting from an arbitrary initial point $x_0\in E$, the iterates very quickly converge on a much smaller subset, a quasi-attractor $Q$. The dynamics on $Q$ is chaotic, and $Q$ would be a true (chaotic) attractor in an ill-posed problem instance, when $A\cap B$ is empty. Since the two projections are in fact very insensitive to the existence of a solution, it follows that the dynamics in a well-posed instance is similar, only differing when the iterate arrives at the attractive basin of a fixed point and the algorithm terminates. A cartoon comparison of exhaustive ``rotamer" search and difference map search is given in Figure 3. 

\begin{figure}[t]
\begin{center}\includegraphics[width=3.5in]{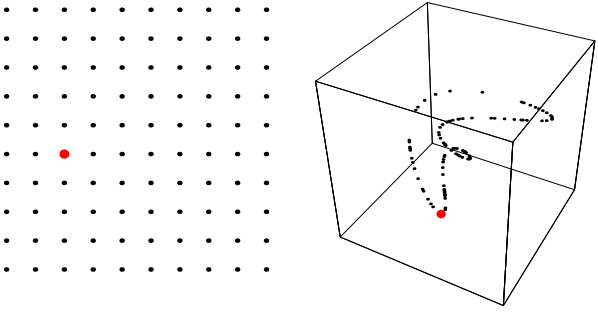}
\end{center}
\begin{center}
\caption{Spaces sampled by optimization algorithms: ``rotamer" space for two dihedral angles (left), difference map quasi-attractor (right). The dimension of the quasi-attractor is smaller than that of the rotamer space, even though it is embedded in a higher dimensional Euclidean space. The large point represents the solution.}
\end{center}
\end{figure}

\textbf{Heteropolymer models.} We consider two off-lattice heteropolymer models, with monomer-monomer interaction of the Lennard-Jones form:
\begin{equation}\label{LJ}
E_\mathrm{LJ}=4 \sum_{i=1}^{N-2}\sum_{j=i+2}^{N}\left(\frac{1}{r_{i j}^{12}}-\frac{C_{i j}}{r_{i j}^6}\right)\; .
\end{equation}
$N$ is the number of monomers, $\mathbf{r}_{i j}$ is the vector separation of monomers $i$ and $j$ with magnitude $|\mathbf{r}_{i j}|=r_{i j}$, and $C_{i j}=C_{j i}$ are constants that depend on the hydrophobic (H) and polar (P) character of the monomers. For the \textit{flexible chain} model proposed by Stillinger \textit{et al.}\cite{Stillinger},
\begin{equation}
C_\mathrm{HH}=1\qquad C_\mathrm{HP}=-\frac{1}{2}\qquad C_\mathrm{PP}=\frac{1}{2}\;.
\end{equation}
Another model we study, the \textit{rotamer} model, has
\begin{equation}
C_\mathrm{HH}=1\qquad C_\mathrm{HP}=C_\mathrm{PP}=\frac{1}{2}\;.
\end{equation}
The main difference between the flexible chain and rotamer models is the nature of the constraints on the polymer chain. In the flexible chain model only the bond length is fixed, $r_{i \,i+1}=1$; in the rotamer model the bond angles are fixed as well:
$\mathbf{r}_{i-1 \,i}\cdot \mathbf{r}_{i \,i+1}=\cos{\alpha}$. Since the latter constraint fixes the distances $r_{i\, i+2}$, these terms are excluded from the sum in (\ref{LJ}) for the rotamer model. The flexible chain model adds a bond angle energy favoring linear conformations:
\begin{equation}\label{chain}
E_\mathrm{chain}=\frac{1}{4}\sum_{i=2}^{N-1}(1-\mathbf{r}_{i-1 \,i}\cdot \mathbf{r}_{i \,i+1})\; .
\end{equation}

\textbf{Constraint projections.} Protein conformations are subject to two, typically antagonistic, constraints. In order to function, proteins adopt a compact shape with stability and functionality conferred by the three dimensional packing of its constituent amino acid residues. In order for the protein to be synthesized, however, the arrangement of the residues must also correspond to a possible conformation of a polypeptide chain. Either of these constraints would be much easier to satisfy if the other could be neglected, and there would then be a multitude of solutions. The difficulty in finding the native fold, from this perspective, is finding a configuration of residues that satisfies both constraints. We discuss later how this point of view provides a basis for understanding the uniqueness of the native fold.

The application of the difference map algorithm to the model proteins described above involves three things: specifying the embedding, defining the constraint sets, and computing projections to the constraint sets. We embed both models in a Euclidean space $E$ of dimension $3N$ in the standard way: three Cartesian coordinates for each monomer position. The constraint sets $A$ and $B$ correspond to the chain constraints and packing constraints, respectively. 

Set $A$ in the flexible chain model is the set of all monomer configurations in $E$ with $r_{i \,i+1}=1$, while in the rotamer model we impose the additional constraint $\mathbf{r}_{i-1 \,i}\cdot \mathbf{r}_{i \,i+1}=\cos{\alpha}$ (for a given $\alpha$). The projection to $A$, or $P_A$, is computed with the aid of a penalty function $V_\mathrm{chain}$. For the rotamer model we used
\begin{equation}\label{penalty}
V_\mathrm{chain}=\sum_{i=1}^{N-1}(r_{i\, i+1}-1)^2+\sum_{i=2}^{N-1}(\mathbf{r}_{i-1 \,i}\cdot \mathbf{r}_{i \,i+1}-\cos{\alpha})^2\; .
\end{equation}
The flexible chain model used only the first term in (\ref{penalty}). To compute $P_A(x)$, given some input monomer configuration $x\in E$, we use gradient descent minimization of $V_\mathrm{chain}$, terminated when the step size falls below a given threshold. The algorithm records the success of the projection by testing whether $V_\mathrm{chain}$ is within a small tolerance value of zero. In the experiments reported below, the success rate for $P_A$ was 100\%.

\begin{figure}[t]
\begin{center}\includegraphics[width=2.5in]{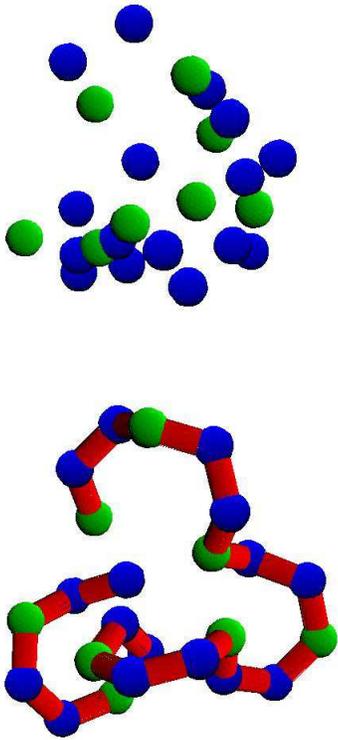}
\end{center}
\begin{center}
\caption{Chain constraint projection applied to a monomer configuration (top) in the rotamer model.}
\end{center}
\end{figure}

The packing constraint set $B$ in the rotamer model is simply the set of monomer configurations $x\in E$ satisfying $E_\mathrm{LJ}(x)<E_0$, where $E_0$ specifies the energy depth of the search. For the constraint satisfaction problem to have a feasible point, or the difference map to have a fixed point, $E_0$ must be greater than the ground state energy of the polymer. We again compute the corresponding projection, $P_B$, using gradient descent, but now with the function $E_\mathrm{LJ}$. The termination criterion is also different, since we are only interested in crossing the $E_\mathrm{LJ}(x)=E_0$ contour, rather than finding a local minimum. After crossing the target contour, we use Newton iterations to converge on the contour. In the event that the input $x$ already satisfies $E_\mathrm{LJ}(x)<E_0$, the same $x$ is returned as the output of the projection.  Crossing of the $E_0$ contour is used as the criterion for a successful computation of $P_B$. Clearly the success rate depends on $E_0$. In our experiments the success rate for $P_B$ was essentially 100\%, since the target energy $E_0$ is always such that finding a feasible point of $E_\mathrm{LJ}(x)<E_0$ is easy. This is because the target energies of relevance, those that apply in the dual constraint problem, are always significantly above the minimum energy of the pure packing problem (no chain constraint). Because the inputs to projections generated by the difference map scheme can fall within regions where $E_\mathrm{LJ}$ diverges sharply, we modified the Lennard-Jones potential to have the form $a-b\,r_{i j}^2$ for separations $r_{i j}<0.9$, with $a$ and $b$ chosen to make $E_\mathrm{LJ}$ and its first derivatives continuous. All the folds discovered by the algorithm have $r_{i j}>0.9$ for all monomer pairs $i$ and $j$.

Figure 4 shows the action of the chain constraint projection, $P_A$, on a configuration of monomers in the rotamer model with $\cos{\alpha}=0.5$. The H/P sequence is known only to $P_A$; in this example it is periodic with a three element motif: $(\mathrm{HPP})_8$. The packing constraint, $P_B$, is blind to the sequence ordering of monomers.

The formulation of the packing constraint set, and the computation of its projection, was somewhat different in the flexible chain model. This example illustrates both the pitfalls in the naive application of the difference map algorithm, as well as its flexibility. The chain energy (\ref{chain}) would seem to have its natural place in defining the chain constraint $A$. However, this would entail having to specify another adjustable energy parameter in addition to the packing energy $E_0$. The other option, of combining $E_\mathrm{chain}$ with $E_\mathrm{LJ}$ (thereby modifying set $B$), would be a mistake because the former has a very long-range character, in contrast with the latter, and the projection would almost always be blind to the possibility of favorable monomer contacts. Our solution was to combine a modified form of $E_\mathrm{chain}$ with $E_\mathrm{LJ}$:
\begin{equation}
E^\prime_\mathrm{chain}=\frac{1}{4}\sum_{i=2}^{N-1}(1-\mathbf{r}_{i-1 \,i}\cdot \mathbf{r}_{i \,i+1})w(r_{i-1 \,i})w(r_{i \,i+1})\; ,
\end{equation}
where
\begin{equation}
w(r)=\left\{
\begin{array}{ll}
1&\mbox{if $r\le1$}\\
1-(1/r^2-1)^2&\mbox{if $r>1$ .}
\end{array}\right.
\end{equation}
A modification of this kind is valid, since any solution $x\in A\cap B$ satisfies the chain constraint, and $E^\prime_\mathrm{chain}$ reduces to $E_\mathrm{chain}$.

Gradient descent to a constraint set specified by the contours of a function, is only distance minimizing when the constraint function is linear. We considered the possibility, when seeking a nearby point on a contour, say $V(x)=V_0$, that it may be advantageous to perform gradient descent on a ``guiding function", say $G(x)$. The descent would still be terminated at the contour of the original function; the role of the guiding function is only to minimize the length of the path to the contour. In the rotamer model we obtained good quality projections without the use of guiding functions. In the flexible chain model, however, we used the guiding function
\begin{equation}\label{guide}
G(x)=E_\mathrm{LJ}(C_\mathrm{HP}; x)\;.
\end{equation}
$G(x)$ omits the chain bending energy $E^\prime_\mathrm{chain}$ and allows for a modified value of the Lennard-Jones parameter $C_\mathrm{HP}$. The negative value of $C_\mathrm{HP}$ in the model has the effect that during gradient descent the condensed monomers may fission into separated H and P domains. This is avoided by giving $C_\mathrm{HP}$ a non-negative value in the guiding function.

\section*{Results}

\textbf{Rotamer model.} A useful record of the progress of the difference map algorithm is the time series of difference magnitudes, $\delta_t=\|\Delta(x_t)\|$. In our folding application, $\delta_t$ is the rms-displacement (in units of the chain's bond length) of monomers in two configurations: one satisfying the chain constraint, the other satisfying the packing (energy) constraint. The algorithm terminates when $\delta_t = 0$, that is, when a valid polymer geometry is found with energy below the chosen target value $E_0$ (a ``feasible solution"). Figure 5 shows a difference plot with $\beta=1.2$ for the sequence $(\mathrm{HPP})_8$ in the rotamer model with geometry $\cos{\alpha}=0.5$ and target energy $E_0=-24$. The behavior of $\delta_t$ in the folding problem is typical of behavior observed in other applications\cite{diffmap}. The three stages of the solution process are evident in (1) the initial (very fast) decay during convergence to the quasi-attractor, (2) steady-state fluctuations as the quasi-attractor is searched, and (3) a final (fast) decay to zero when the solution (a fixed point) is discovered. As in phase retrieval, the distribution of run times (total iterations) is exponential\cite{retriever} and consistent with the interpretation of a very fast relaxation of the probability distribution on the quasi-attractor. For the parameters given, the average number of iterations per solution was $I_\mathrm{ave}=7500$.

The feasible solutions found by the difference map for given target energies $E_0$ were refined by steepest descent minimization of the heteropolymer energy; the chain geometry was maintained by adding the penalty function (\ref{penalty}) with a large multiplier. For each run of the algorithm we therefore obtain one locally minimized fold with energy guaranteed to be below $E_0$. In the example above, about half of the outputs had the same refined energy of $-25.048$ and  structure (or enantiomorph). Since this also is the lowest energy obtained, we have good reason to believe this is the ground state. The structure, shown in Figure 6, resembles a cut trefoil knot.

\begin{figure}[t]
\begin{center}\includegraphics[width=3in]{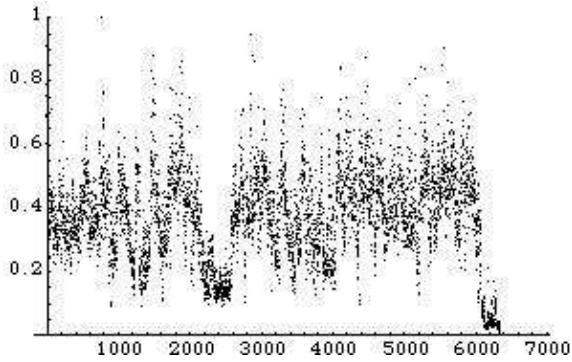}
\end{center}
\begin{center}
\caption{Evolution of the rms displacement of monomers, $\delta_t$, between two configurations that satisfy the chain and packing constraints, respectively. The fold shown in Figure 6 was found in just over $6000$ iterations.}
\end{center}
\end{figure}

\begin{figure}[t]
\begin{center}\includegraphics[width=2in]{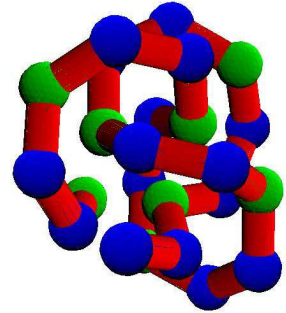}
\end{center}
\begin{center}
\caption{The fold having the lowest energy for the sequence $(\mathrm{HPP})_8$ in the rotamer model has the shape of a cut trefoil knot.}
\end{center}
\end{figure}

The most direct measure of the work performed by the algorithm is the average number of iterations per solution $I_\mathrm{ave}$, divided by the rate $p_0$ with which the lowest energy fold (putative ground state) is obtained. This is a number that we expect to grow exponentially with the length of the polymer, and roughly corresponds to the number of conformations that must be sampled before one can claim to have discovered the ground state. For the example above, $I_\mathrm{ave}/p_0\approx 15000$. We repeated the above experiment with longer sequences having the same repeating motif. The size $N=36$ is about the limit of where the ground state can be established with modest computing resources (a single processor). As argued below, it may be possible to exceed this limit for ``well designed" sequences. Our rotamer model experiments are summarized in Table 1.

\textbf{Flexible chain model.} Studies of this model by other investigators\cite{Hsu,Bachmann} have been limited to Fibonacci sequences $F_k$, defined by
\begin{equation}
F_0=\mathrm{H}, \;F_1=\mathrm{P}\;, \qquad F_{k+1}=F_{k-1}F_k\;.
\end{equation}
The tendency toward hydrophobic core formation is even stronger for the Lennard-Jones parameters of the flexible chain model. For the Fibonacci sequences, in particular, the chain bending energy must be sacrificed in order to allow the chain to weave between the hydrophobic core and polar envelope. To improve the packing projection we therefore used the guiding function (\ref{guide}), which omits the bending energy, and $C_\mathrm{HP}=0$ for the short chains, $C_\mathrm{HP}=0.1$ for $N=55$. A sign change of the difference map parameter $\beta$, which effectively interchanges the two constraint sets, gave somewhat better results in the flexible chain model.

Our results for Fibonacci chains up to $N=55$ are summarized and compared with other algorithms in \mbox{Table 2}. The difference map corroborates the ground state candidates found by the ELP\cite{elp} algorithm for chains up to $N=34$, and finds a lower energy fold for $N=55$. All the best folds have a well developed hydrophobic core; the $N=55$ chain shown in Figure 7 is a good example. The latter fold was only obtained in one run, and we are therefore far from claiming to have found the ground state. 

\begin{figure}[t]
\begin{center}\includegraphics[width=3.5in]{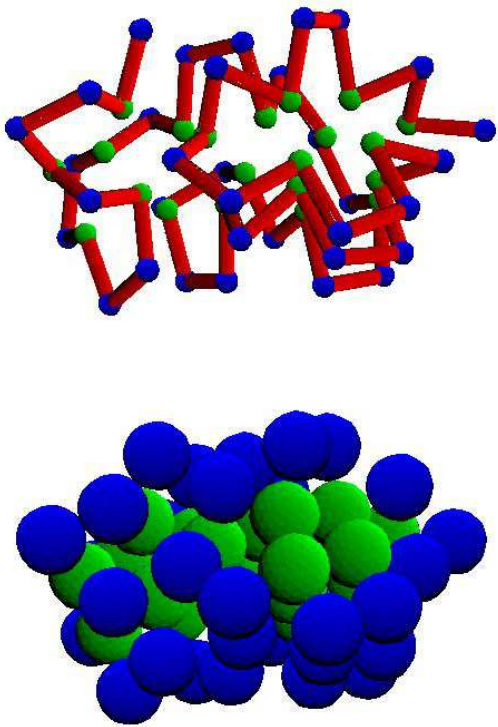}
\end{center}
\begin{center}
\caption{Fold with lowest known energy for the $N=55$ Fibonacci sequence in the flexible chain model; \textit{top:} chain geometry, \textit{bottom:} monomer packing.}
\end{center}
\end{figure}

\begin{figure}[t]
\begin{center}\includegraphics[width=3.5in]{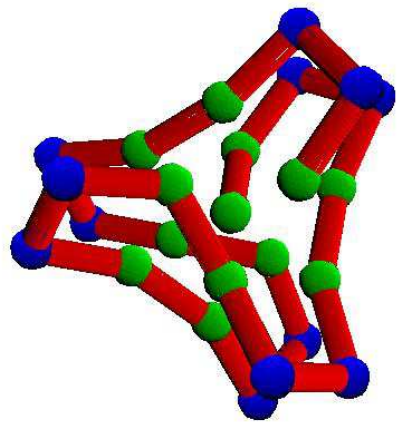}
\end{center}
\begin{center}
\caption{Ground state of the designed sequence $\mathrm{H}(\mathrm{HPPH})_6$ in the flexible chain model. The 13 H monomers form the vertices and center of an almost perfect icosahedron. Apart from the final bond in the structure, the chain geometry is approximately symmetric with respect to a 2-fold axis.}
\end{center}
\end{figure}

Low energy folds in the flexible chain model for sequences containing adjacent H monomers are qualitatively different from the low energy folds for Fibonacci sequences, in which H monomers are never adjacent. A good example is provided by the $N=25$ sequence $\mathrm{H}(\mathrm{HPPH})_6$, which was designed to realize a particular ground state geometry. Using the difference map it is easy to establish the ground state shown in Figure 8. This fold is unique in that it simultaneously minimizes the bending energy where the chain passes through the icosahedral core, and also arranges the six hairpin turns so that the P mono\-mers there form the largest number of contacts.   

\section*{Discussion}

The difference map folding algorithm was shown to be competitive with leading algorithms in experiments with model proteins. We conclude by discussing two issues that will be important in applications to realistic protein models.

\textbf{Designed sequences.} The performance of an iterative phase retrieval algorithm, of which the difference map folding algorithm is a logical descendent, is sensitively dependent on the degree to which the input data is overdetermined\cite{diffmap}. We believe that the latter attribute's counterpart in protein folding is the property of being ``well designed". 

In the context of the geometry of the difference map, a highly overdetermined problem corresponds to the situation where the probability of nonempty intersection of the constraint sets $A$ and $B$, given a specification by random data, is exceedingly small. This makes the existence of a solution all the more unusual. In phase retrieval one is guaranteed a solution in even these unlikely circumstances, and moreover, the uniqueness of the solution and efficiency of the solution process relies on this fact.

\begin{table*}[t]
\begin{center}
\begin{tabular}{rl|rrrrll}
 $N$ & sequence  & $E_\mathrm{DM}$       & $E_0$  & $\beta$ & $I_\mathrm{ave}$ & $p_0$ & time/iter
  \\
\hline
$24$ & $(\mathrm{HPP})_8$    & $-25.048$      & $-24.0$      & $1.2$     & $7500$ &  $0.50$    &   $12$ msec   \\

$30$ & $(\mathrm{HPP})_{10}$    & $-34.900$      & $-33.0$      & $1.2$     & $23000$ &  $0.07$    &  $18$    \\

$36$	& $(\mathrm{HPP})_{12}$ & $-45.851$	& $-42.5$& $1.2$	& $150000$	&---	&  $26$\\

\hline
\end{tabular}
\caption{Results for the rotamer model. $E_0$ is the target energy of the difference map (DM) algorithm, $I_\mathrm{ave}$ the average number of iterations to find the target energy, and $p_0$ is the probability that the discovered fold refines to the lowest energy obtained in the experiment, $E_\mathrm{DM}$. The last column gives the cpu time per iteration on a 1.67 GHz processor.}
\end{center}
\end{table*}

\begin{table*}[t]
\begin{center}
\begin{tabular}{rc|rrr|rrrrll}      
 $N$ & sequence    & $E_\mathrm{PERM}$    & $E_\mathrm{MUCA}$       & $E_\mathrm{ELP}$	& $E_\mathrm{DM}$       & $E_0$ & $\beta$ & $I_\mathrm{ave}$ & $p_0$ & time/iter
  \\
\hline
$13$ & $F_6$     & $-4.962$      & $-4.967$      & $-4.967$     &  $-4.975$    &  $-4.5$    & $-1$ & $34$ & $0.34$  & $3$ msec\\

$21$	& $F_7$ & $-11.524$	& $-12.296$	& $-12.316$	&$-12.327$	& $-11.8$ & $-1$	& $2900$	& $0.024$ & $25$ \\

$34$	& $F_8$ & $-21.568$	& $-25.321$	& $-25.476$	&$-25.512$	& $-23.5$	&$-1$  &$10000$	& $0.007$ &	$80$\\

$55$   & $F_9$   & $-32.884$      & $-41.502$      & $-42.428$     & $-43.331$     & $-38.0$     & $-1$  & $27000$  & ---  & $200$\\

$25$   & $\mathrm{H}(\mathrm{HPPH})_6$   &       &     &      &$-28.313$      & $-27.4$   & $-1$  & $9200$   &$0.030$   & $45$\\
\hline
\end{tabular}
\caption{Results for the flexible chain model. Ground state energy estimates obtained by the difference map (DM) algorithm are compared with three other algorithms: pruned-enriched Rosenbluth method (PERM\cite{Hsu}), multicanonical sampling (MUCA\cite{Bachmann}), and energy landscape paving (ELP\cite{Bachmann}). Optimal structures found by ELP and DM for $N=13, 21$, and $34$ Fibonacci sequences are essentially the same\cite{BachmannPrivate}. For $N=55$ DM finds a different, lower energy fold. See Table 1 for definitions of DM parameters.}
\end{center}
\end{table*}

Whether the simple protein models studied above have the capacity for realizing highly overdetermined problem instances (sequences) is open to speculation. With our choice of deconstructing the energy landscape into chain and packing constraints, this would imply the existence of exceptionally low energy monomer packings that nevertheless can be threaded by a particular sequence. Folds with these properties should be easier to find, because the target energy $E_0$ of the difference map algorithm could be set at a lower value and thereby eliminate a large part of the energy landscape. One experiment to test this hypothesis, in the flexible chain model, would be to fold random sequences of 13 H and 12 P monomers and compare performance, as well as ground state energies, with the designed sequence $\mathrm{H}(\mathrm{HPPH})_6$.

\textbf{More realistic models.} A serious deficiency in the protein models studied above is the omission of  the hydrogen bonding mechanism that acts on the peptide geometry and is responsible for the two distinctive types of secondary structure. Implementing this level of detail lies at the heart of applying the difference map algorithm to realistic models. How big a space should the constraints be embedded within? It is cetainly not enough, as in the rotamer and flexible chain models, to embed a protein of $N$ residues in a space of dimension $3 N$. The orientations and internal dihedral angles of side groups require additional coordinates in their specification. The peptide chain geometry is also more complicated, and with its two dihedral angles per residue may require twice the embedding dimension than its analog in the rotamer model. 

These remarks should make clear that there is no automatic procedure for effectively applying the difference map algorithm to arbitrary optimization problems. An essential part of the endeavor is the formulation as a constraint satisfaction problem, that is, a recipe for deconstructing the energy landscape.

\section*{Acknowledgements}
Veit Elser thanks Ron Elber for discussions. This work was supported by grant NSF-DMR-0426568 and GAANN award P200A030111 from the U.S. Dept. of Education.

\end{document}